# The strain-induced transitions of the piezoelectric, pyroelectric and electrocaloric properties of the CuInP$_2$S$_6$ films

Anna N. Morozovska[1*], Eugene A. Eliseev[2], Lesya P. Yurchenko[2], Valentyn V. Laguta[2,3], Yongtao Liu[4], Sergei V. Kalinin[5], Andrei L Kholkin[6], and Yulian M. Vysochanskii[7]

[1] Institute of Physics, National Academy of Sciences of Ukraine,
46, Prospekt Nauky, 03028 Kyiv, Ukraine,

[2] Frantsevich Institute for Problems of Materials Science, National Academy of Sciences of Ukraine, 3, Omeliana Pritsaka, 03142 Kyiv, Ukraine,

[3] Institute of Physics, Czech Academy of Sciences, 18200 Prague, Czech Republic

[4] Center for Nanophase Materials Sciences, Oak Ridge National Laboratory,
Oak Ridge, TN 37922, United States,

[5] Department of Materials Science and Engineering, University of Tennessee,
Knoxville, TN, 37996, United States,

[6] Department of Physics & CICECO – Aveiro Institute of Materials, University of Aveiro, 3810-193 Aveiro, Portugal,

[7] Institute of Solid-State Physics and Chemistry, Uzhhorod University,
88000 Uzhhorod, Ukraine.

## Abstract

The low-dimensional ferroelectrics, ferrielectrics and antiferroelectrics are of urgent scientific interest due to their unusual polar, piezoelectric, electrocaloric and pyroelectric properties. The strain engineering and strain control of the ferroelectric properties of layered two-dimensional Van der Waals materials, such as CuInP$_2$(S,Se)$_6$ monolayers, thin films and nanoflakes, are of fundamental interest and especially promising for their advanced applications in nanoscale nonvolatile memories, energy conversion and storage, nano-coolers and sensors. Here, we study the polar, piezoelectric, electrocaloric and pyroelectric properties of thin strained films of a ferrielectric CuInP$_2$S$_6$ covered by semiconducting electrodes and reveal an unusually strong effect of a mismatch strain on these properties. In particular, the sign of the mismatch strain and its magnitude determine the complicated behavior of piezoelectric, electrocaloric and pyroelectric responses. The strain effect on these properties is opposite, i.e., "anomalous", in comparison with many other ferroelectric films, for which the out-of-plane remanent polarization, piezoelectric, electrocaloric and pyroelectric responses increase strongly for tensile strains and decrease or vanish for compressive strains.

---

[*] Corresponding author: anna.n.morozovska@gmail.com

## I. INTRODUCTION

The piezoelectric, pyroelectric and electrocaloric effects are inherent to ferroelectrics, being a consequence of their spontaneous polarization dependence on strain and temperature, which becomes especially strong in the vicinity of the paraelectric - ferroelectric phase transition. These properties determine the indispensable value of ferroelectrics for modern actuators, pyroelectric sensors, electromechanical and electrocaloric energy converters [1, 2].

The low-dimensional ferroelectrics, ferrielectrics and antiferroelectrics are of urgent scientific interest due to their unusual polar, piezoelectric, electrocaloric and pyroelectric properties [3, 4]. The strain engineering and strain control of the ferroelectric properties of layered two-dimensional Van der Waals (**V-d-W**) materials, such as $CuInP_2(S,Se)_6$ monolayers, thin films and nanoflakes, are of fundamental interest and especially promising for their advanced applications in nanoscale nonvolatile memories, energy conversion and storage, nano-coolers and sensors [5].

One of the most important feature, which determine the strain-polarization coupling in ferrielectric $CuInP_2(S,Se)_6$ [6, 7], is the existence of more than two potential wells [8], which are responsible for strain-tunable multiple polar states [9]. Due to the multiple potential wells, which height and position are temperature- and strain-dependent, the energy profiles of a uniaxial ferrielectric $CuInP_2(S,Se)_6$ can be flat in the vicinity of the nonzero polarization states. The flat energy profiles give rise to the unusual polar and dielectric properties associated with the strain-polarization coupling in the vicinity of the states [10, 11, 12].

The spontaneous polarization of crystalline $CuInP_2S_6$ (**CIPS**) is directed normally to its structural layers being a result of antiparallel shifts of the $Cu^+$ and $In^{3+}$ cations from the middle of the layers [13, 14]. The strain effect on the polarization reversal in CIPS is opposite, i.e., "anomalous", in comparison with many other ferroelectric films, for which the out-of-plane remanent polarization and coercive field increase strongly for tensile strains, and decrease or vanish for compressive strains [9 - 12].

Using the Landau-Ginzburg-Devonshire (**LGD**) approach here we study the size- and strain-induced changes of the spontaneous polarization, piezoelectric, pyroelectric and electrocaloric properties of thin strained CIPS films covered by semiconducting electrodes. The original part of this work contains the physical description of the problem (**Section II**), analysis the strain-induced transitions of the piezoelectric, electrocaloric and pyroelectric properties of the CIPS films (**Section III**). **Section IV** summarizes the obtained results. **Supplementary Materials** elaborate on a mathematical formulation of the problem and the table of material parameters.

## II. PROBLEM FORMULATION

Let us consider an epitaxial thin CIPS film sandwiched between the semiconducting electrodes with a screening length λ, which is clamped on a thick rigid substrate [see **Fig. 1(a)**]. Arrows show the out-of-plane ferroelectric polarization $P_3$, directed along the $X_3$-axis. The perfect electric contact between the film and the electrodes provides the effective screening of the out-of-plane polarization by the electrodes and precludes the domain formation for small enough λ. An electric voltage is applied between the electrodes. The misfit strain $u_m$ originates from the film-substrate lattice constants mismatch and exists entire the film depth [15, 16, 17], because the film thickness $h$ is regarded smaller than the critical thickness $h_d$ of misfit dislocations appearance.

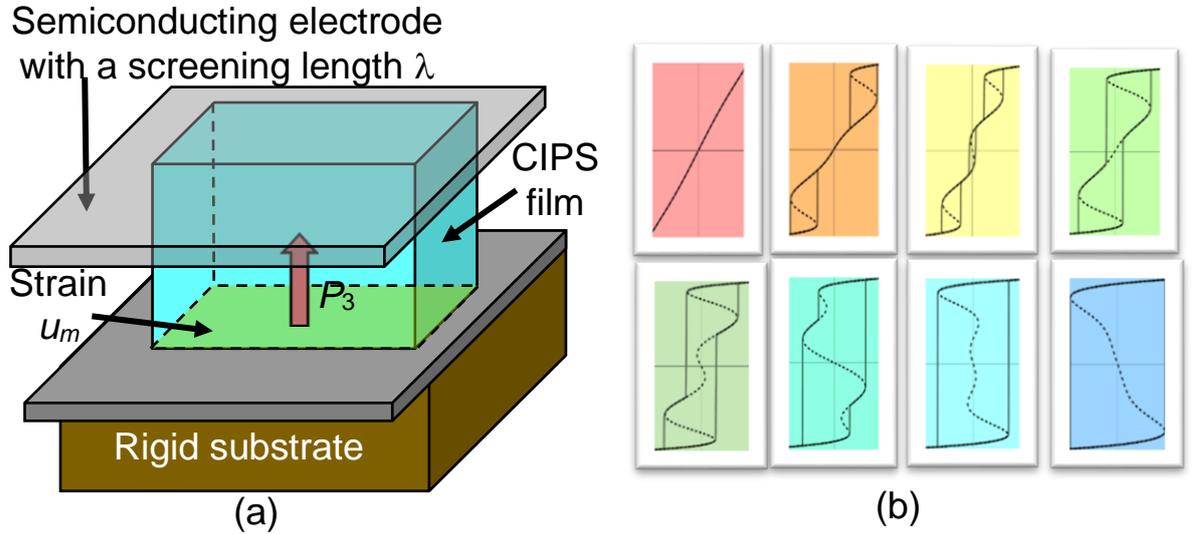

**FIGURE 1**. **(a)** Schematics of a thin epitaxial CIPS film sandwiched between semiconducting electrodes with a small screening length λ, and clamped on a rigid substrate. Arrow shows the direction of the single-domain spontaneous polarization. **(b)** Schematics of the possible strain-induced changes of polarization reversal hysteresis loops.

Within the LGD approach, the value and orientation of the spontaneous polarization $P_i$ in thin ferroelectric films are controlled by the temperature $T$ and mismatch strain $u_m$. For the validity of the continuum media approximation, the film thickness is regarded to be much bigger than the lattice constant $c$. As a rule, the condition $c \ll h < h_d$ is valid for the film thickness range (5 – 50) nm.

It has been shown in Refs.[9 - 12] that the LGD free energy density of CIPS, $g_{LGD}$, has four potential wells at $\vec{E} = 0$. The density $g_{LGD}$ includes the Landau-Devonshire expansion in even powers of the polarization $P_3$ (up to the eighth power), the Ginzburg gradient energy, the elastic and electrostriction energies. The behavior of polarization $P_3$, piezocoefficient $d_{33}$, and pyroelectric coefficient $\Pi_3$ in the electric field $E_3$ follows from the time-dependent Euler-Lagrange equations, which have the form [9 - 10]:

$$\Gamma\frac{\partial P_3}{\partial t} + \left[\alpha - 2\sigma_i(Q_{i3} + W_{ij3}\sigma_j)\right]P_3 + (\beta - 4Z_{i33}\sigma_i)P_3^3 + \gamma P_3^5 + \delta P_3^7 - g_{33kl}\frac{\partial^2 P_3}{\partial x_k \partial x_l} = E_3, \quad (1a)$$

$$\Gamma\frac{\partial d_{33}}{\partial t} + \left[\alpha - 2\sigma_i(Q_{i3} + W_{ij3}\sigma_j) + 3(\beta - 4Z_{i33}\sigma_i)P_3^2 + 5\gamma P_3^4 + 7\delta P_3^6\right]d_{33} - g_{33kl}\frac{\partial^2 d_{33}}{\partial x_k \partial x_l} =$$
$$(2Q_{33} + 2W_{i33}\sigma_i)P_3 + 4Z_{333}P_3^3. \quad (1b)$$

$$\Gamma\frac{\partial \Pi_3}{\partial t} + \left[\alpha - 2\sigma_i(Q_{i3} + W_{ij3}\sigma_j) + 3(\beta - 4Z_{i33}\sigma_i)P_3^2 + 5\gamma P_3^4 + 7\delta P_3^6\right]\Pi_3 - g_{33kl}\frac{\partial^2 d_{33}}{\partial x_k \partial x_l} =$$
$$\alpha_T P_3 + \beta_T P_3^3 + \gamma_T P_3^5. \quad (1c)$$

Here, $\Gamma$ is the Khalatnikov kinetic coefficient [18]. The coefficient $\alpha$ depends linearly on the temperature $T$, namely $\alpha(T) = \alpha_T(T - T_C)$, where $T_C$ is the Curie temperature of a bulk ferrielectric. The coefficients $\beta$, $\gamma$, and $\delta$ are temperature independent. The values $\sigma_i$ denote diagonal components of a stress tensor in the Voigt notation, and the subscripts $i$ and $j$ vary from 1 to 6. The values $Q_{i3}$, $Z_{i33}$, and $W_{ij3}$ denote the components of a second order and higher order electrostriction strain tensors in the Voigt notation, respectively [19, 20]. The values $g_{33kl}$ are polarization gradient coefficients in the matrix notation and the subscripts $k, l = 1 - 3$. The boundary condition for $P_3$ at the film surfaces S is regarded "natural", i.e., $g_{33kl}n_k\frac{\partial P_3}{\partial x_l}\Big|_S = 0$, where $\vec{n}$ is the outer normal to the surface.

The values of $T_C$, $\alpha_T$, $\beta$, $\gamma$, $\delta$, $Q_{i3}$, $W_{ij3}$, and $Z_{i33}$ have been derived in Ref.[21] from the fitting of temperature-dependent experimental data for the dielectric permittivity [22, 23], spontaneous polarization [24], and lattice constants [25] as a function of hydrostatic pressure. Elastic compliances $s_{ij}$ were calculated from ultrasound velocity measurements [26, 27]. The CIPS parameters are listed in **Table SI** in **Appendix A** (see Supplementary materials).

Modified Hooke's law, relating elastic strains $u_i$ and stresses $\sigma_j$, is obtained from the relation $u_i = -\partial g_{LGD}/\partial \sigma_i$:

$$u_i = s_{ij}\sigma_j + Q_{i3}P_3^2 + Z_{i33}P_3^4 + W_{ij3}\sigma_j P_3^2. \quad (2)$$

For the considered geometry of a CIPS film the following relations are valid for homogeneous stress and strain components: $\sigma_3 = \sigma_4 = \sigma_5 = \sigma_6 = 0$, $u_1 = u_2 = u_m$ and $u_4 = u_5 = u_6 = 0$.

The value $E_3$ in Eq.(1a) is an electric field component co-directed with the polarization $P_3$. $E_3$ is a superposition of external ($E_0$) and depolarization ($E_d$) fields. In the considered case of a very high screening degree by the semiconducting electrodes with a small screening length $\lambda \leq 0.1$ nm, the solutions, corresponding to the almost constant $P_3$, are energetically favorable and the domain formation is absent, because the corresponding depolarization field, $E_d = \frac{-P_3}{\varepsilon_0(\varepsilon_b + h/\lambda)}$, is very small for $h/\lambda \gg 1$. To analyze a quasi-static polarization reversal, we assume that the period, $2\pi/\omega$, of the sinusoidal external field $E_0$ is very small in comparison with the Landau-Khalatnikov relaxation time, $\tau = \Gamma/|\alpha|$.

The electrocaloric (**EC**) temperature change $\Delta T_{EC}$, can be calculated from the expression [28]:

$$\Delta T_{EC} = -T \int_{E_1}^{E_2} \frac{1}{\rho_P C_P} \left(\frac{\partial P}{\partial T}\right)_E dE \cong T \int_{E_1}^{E_2} \frac{1}{\rho_P C_P} \Pi_3 \, dE, \quad (3)$$

where $\rho_P$ is the volume density, $T$ is the ambient temperature, and $C_P$ is the CIPS specific heat.

For ferroics the specific heat depends on polarization (and so on external field) and can be modeled as following:

$$C_P = C_P^0 - T \frac{\partial^2 g}{\partial T^2}, \quad (4)$$

where $C_P^0$ is the polarization-independent part of specific heat and $g$ is the density of the LGD free energy. According to experiment, the specific heat usually has a maximum in the first order ferroelectric phase transition point, which height is about (10 – 30) % of the $C_P$ value near $T_C$ (see e.g. [29]). The mass density and the polarization-independent part of the CIPS specific heat are $\rho_P = 3.415 \cdot 10^3$ kg/m$^3$ and $C_P^0 = 3.40 \cdot 10^2$ J/(kg K) [30, 31], respectively.

## III RESULTS AND DISCUSSION

The out-of-plane spontaneous polarization, $P_s$, piezoelectric coefficient, $d_{33}$, pyroelectric coefficient, $\Pi_s$, and electrocaloric temperature change, $\Delta T_{EC}$, as a function of the film thickness $h$ and misfit strain $u_m$, are shown in **Figs. 2**, **3** and **4** for the temperatures $T =$ 250 K, 293 K and 330 K, respectively. The abbreviations "PE", "FE1" and "FE2" mean the paraelectric phase, high-polarization and low-polarization ferrielectric states, respectively.

The color maps of the spontaneous polarization, shown in **Figs. 2(a)**, **3(a)** and **4(a)**, are calculated by a conventional numerical minimization of the free energy (S.1) listed in **Appendix B** (see Supplementary materials). The color scale in the maps shows the absolute value of $P_s$ in µC/cm$^2$ calculated in the deepest potential well of the LGD free energy. A sharp wedge-like region of the PE phase, which is stable at small thickness $h < 10$ nm and small strains $|u_m| < 1\%$, separates two ferrielectric states, FI1 and FI2, which correspond to big and small amplitudes of the out-of-plane spontaneous polarizations $P_3^{\pm}$, respectively. The area and height of the PE phase region significantly increases with the temperature increase, being the smallest for 250 K and the biggest for 330 K. The increase occurs because higher compressive or tensile strains are required to support the FI1 or FI2 polar states, respectively, when the growing temperature increases the coefficient $\alpha(T)$ in Eqs.(1). The unusual features of the color maps are the high-polarization FI1 state existing at compressive strains $u_m < 0$, and the low-polarization state FI2 existing for tensile strains $u_m > 0$. The FI1 and FI2 states either transform into one another for very small $u_m$ or undergoes the first or second order phase transitions to the PE phase. The situation, shown in **Fig. 2** for $u_m > 0$, is anomalous for the most uniaxial and multiaxial ferroelectric films, where the out-of-plane polarization is absent or very small at $u_m > 0$, and the region of the FE c-phase vanishes or significantly constricts for $u_m > 0$ [15].

The color maps of the piezoelectric coefficient, $d_{33}$, shown in **Figs. 2(b), 3(b)** and **4(b)**, are calculated using Eqs.(1a) and (1b) for $E_0 \to 0$. The color scale in the maps shows the absolute value of $d_{33}$ in pm/V. Thin white curves in the plots correspond to the regions where $d_{33}$ diverges at the boundary of the paraelectric-ferrielectric phase transition. The piezoelectric response is smaller in the FI1 state and higher in the FI2 state; and is absent inside the wedge-like region of the PE phase separating the FI states. Relatively small values of $d_{33}$ in the high-polarization FI1 state are explained by the weak field dependence of the saturated out-of-plane spontaneous polarization. Note that $d_{33}$ reaches (60 – 200) pm/V near the PE-FI2 boundary, being the smallest for 330 K and the biggest for 250 K.

The color maps of the pyroelectric coefficient, $\Pi_s$, shown in **Figs. 2(c), 3(c)** and **4(c)**, are calculated using Eqs.(1a) and (1c) for $E_0 \to 0$. The color scale in the maps shows the absolute value of $\Pi_s$ in mC/(K m$^2$). Thin white curves in the plots correspond to the regions where $\Pi_s$ diverges at the boundary of the paraelectric-ferrielectric phase transition. The pyroelectric response is smaller in the FI1 state and significantly higher in the FI2 state; and is absent in the PE phase. Higher values of $\Pi_s$ in the low-polarization FI2 state are explained by the stronger field dependence of the unsaturated out-of-plane low-polarization. Despite the $\Pi_s$ does not exceed 1 mC/(K m$^2$) far from the paraelectric-ferrielectric boundary, the field behavior of $\Pi_s$ determines the features of electrocaloric properties in accordance with Eq.(3).

The color maps of the electrocaloric temperature change, $\Delta T_{EC}$, shown in **Figs. 2(d), 3(d)** and **4(d)**, are calculated using Eqs.(1a), (1c) and (3)-(4) for $E_0 \to E_c$, where $E_c$ is the coercive field of the film. The color scale in the maps shows the value of $\Delta T_{EC}$ in K. The electrocaloric response is smaller in the FI2 state and significantly higher in the FI1 state; and is absent in the PE phase.

Namely, $\Delta T_{EC}$ reaches minimum ~-(2 – 2.5) K near the PE-FI1 boundary for $-1\% < u_m < -0.5\%$ and 40 nm$< h <$10 nm. Higher negative values of $\Delta T_{EC}$ in the high-polarization FI1 state are related with the increase of the spontaneous polarization and with the features of the electrostriction coupling in CIPS, where the second order and higher order coefficients, $Q_{i33}$ and $Z_{i33}$, linearly depend on temperature (see **Table SI** in **Appendix A** from Supplementary materials). Note that $\Delta T_{EC}$ cannot exceed -2.5 K for a bulk BaTiO$_3$ [28], which spontaneous polarization (about 25 µC/cm$^2$ at room temperature) is much higher than the CIPS polarization (about 5 µC/cm$^2$ at room temperature). The negative sign of the electrocaloric effect and its maximum predicted in compressed CIPS films, can be useful for the strain engineering of ultra-thin nano-coolers.

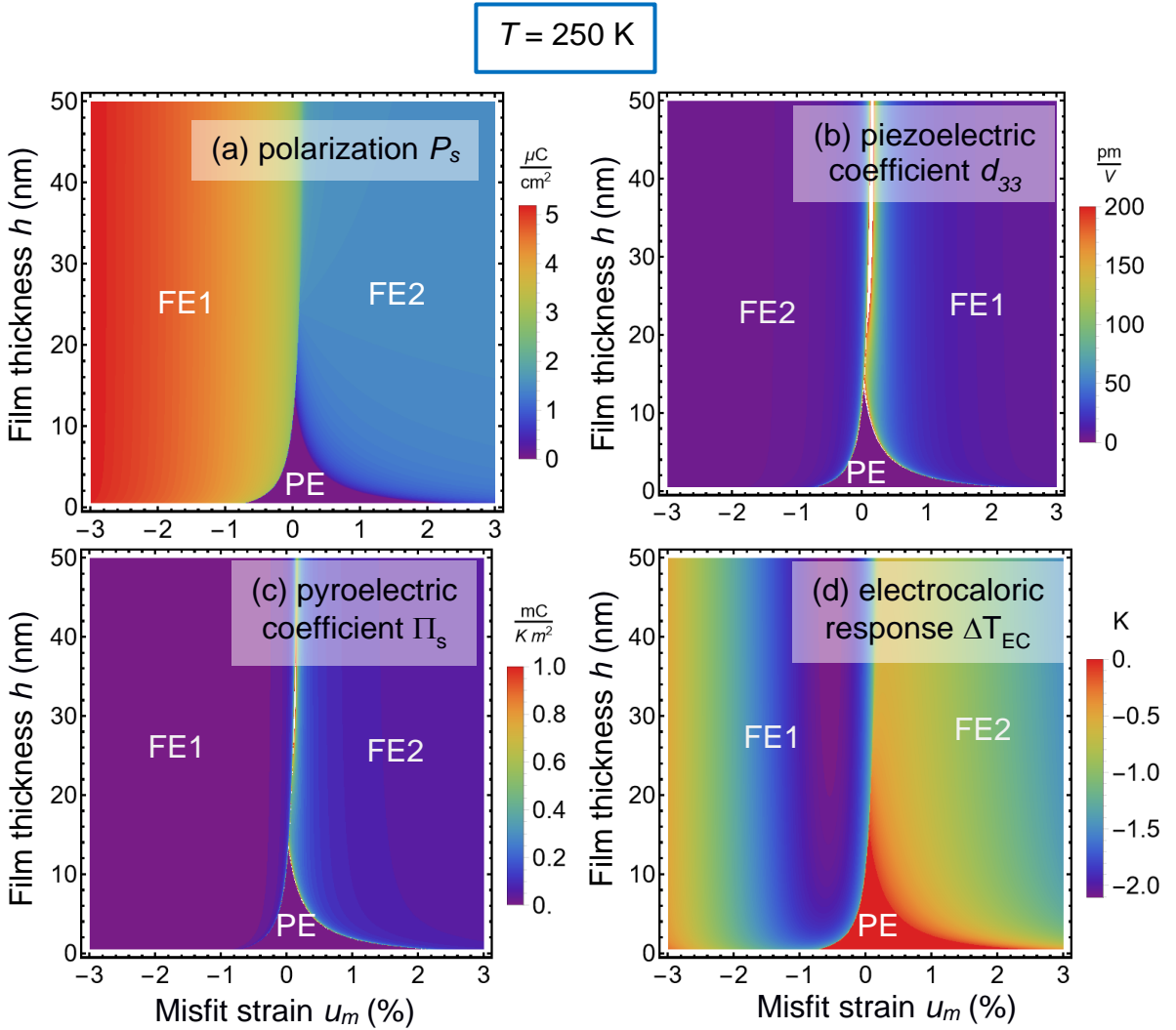

**FIGURE 2.** The spontaneous polarization $P_s$ **(a)**, piezoelectric coefficient $d_{33}$ **(b)**, pyroelectric coefficient $\Pi_s$ **(c)**, and the electrocaloric temperature change $\Delta T_{EC}$ **(d)** of a single-domain CIPS film calculated as a function of the film thickness $h$ and misfit strain $u_m$ for $\lambda = 0.1$ nm and temperature $T =250$ K. Color scales show the values of $P_s$, $d_{33}$, $\Pi_s$ and $\Delta T_{EC}$ in µC/cm², pm/V, mC/(K m²), and K, respectively. White areas in the plots of $d_{33}$ and $\Pi_s$ correspond to the regions, where these values diverge. The abbreviations "PE", "FE1" and "FE2" mean the paraelectric phase, high and low polarization ferrielectric sates, respectively.

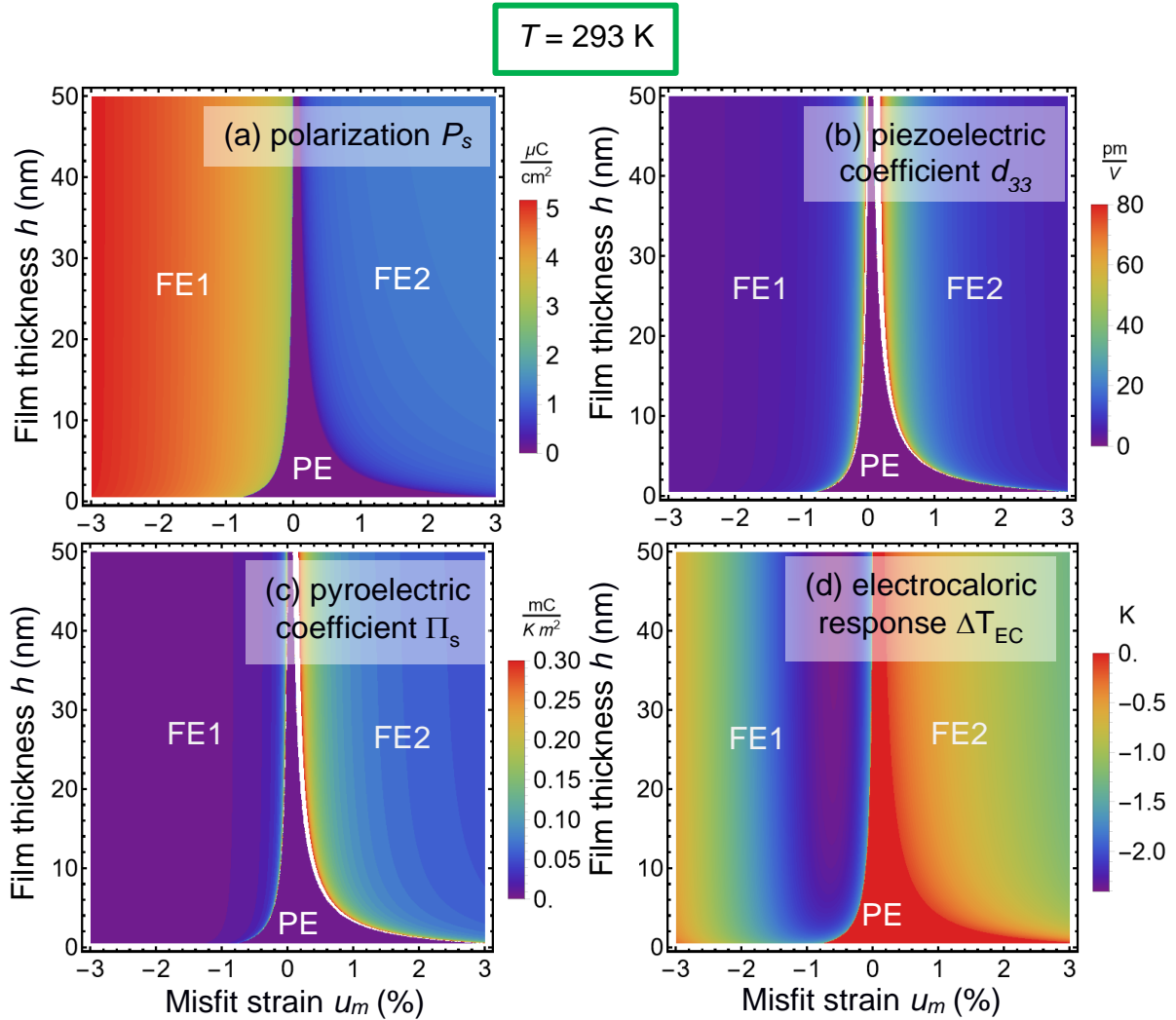

**FIGURE 3.** The spontaneous polarization $P_s$ (**a**), piezoelectric coefficient $d_{33}$ (**b**), pyroelectric coefficient $\Pi_s$ (**c**), and the electrocaloric temperature change $\Delta T_{EC}$ (**d**) of a CIPS film calculated as a function of the film thickness $h$ and misfit strain $u_m$ for the room temperature $T =293$ K. Other parameters and designations the same as in **Fig. 2.**

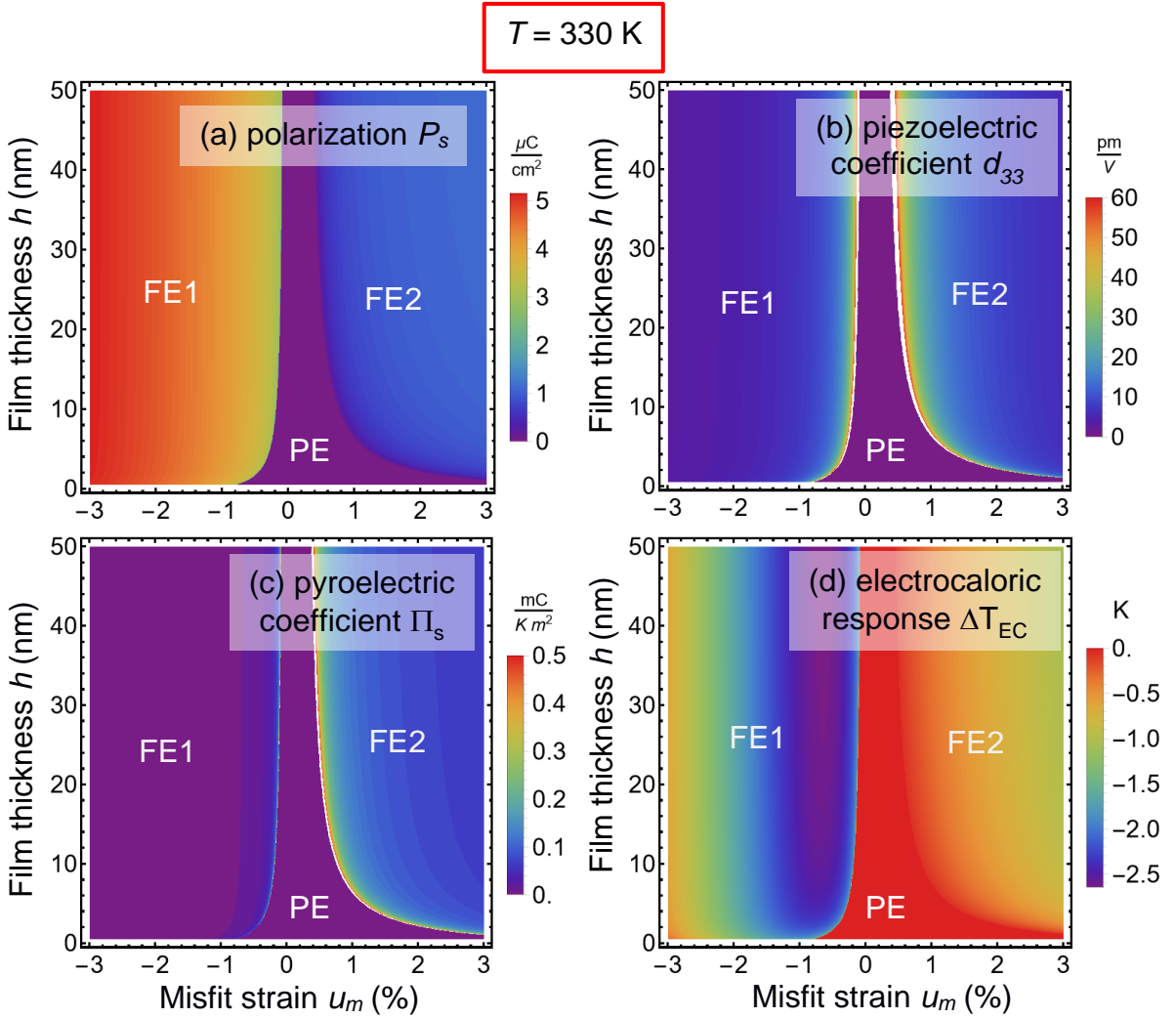

**FIGURE 4.** The spontaneous polarization $P_s$ **(a)**, piezoelectric coefficient $d_{33}$ **(b)**, pyroelectric coefficient $\Pi_s$ **(c)**, and the electrocaloric temperature change $\Delta T_{EC}$ **(d)** of a CIPS film calculated as a function of the film thickness $h$ and misfit strain $u_m$ for the temperature $T = 330$ K. Other parameters and designations the same as in **Fig. 2**.

## IV. SUMMARY

- We consider an epitaxial thin CIPS film sandwiched between semiconducting electrodes, and clamped on a thick rigid substrate, which create the mismatch strain in the film. Using LGD phenomenological approach, we study the piezoelectric, electrocaloric and pyroelectric properties of the strained film and reveal an unusually strong effect of a mismatch strain on these properties.

- We revealed that the sign of the mismatch strain and its magnitude determine the behavior of piezoelectric, electrocaloric and pyroelectric responses. In particular, the strain effect on these properties is opposite, i.e., "anomalous", in comparison with many other ferroelectric films, for which the out-of-plane remanent polarization, piezoelectric, electrocaloric and pyroelectric increase strongly for tensile strains and decrease or vanish for compressive strains. The negative sign of the electrocaloric effect and its maximum predicted in compressed CIPS films, can be useful for the strain engineering of ultra-thin nano-coolers. Also, we studied the conditions when the piezoelectric, electrocaloric and pyroelectric responses can reach maximal values for small electric fields.

**Supplementary materials** contain **APPENDIX A** with LGD parameters for a bulk ferroelectric $CuInP_2S_6$ and **APPENDIX B** with the renormalized free energy caused by misfit stress.

**Author Declaration Section.** The authors have no conflicts to disclose. **Authors' contribution.** **Anna N. Morozovska**: Conceptualization (lead); Writing – review and editing (equal); Formal Analysis (equal), Writing – original draft (lead). **Eugene A. Eliseev**: Software (lead); Visualization (equal); Formal Analysis (equal). **Lesya P. Yurchenko**: Software (supporting). **Valentin V. Laguta**: writing – review and editing (equal). **Yongtao Liu**: review and editing (supporting). **Sergei V. Kalinin**: Writing – review and editing (equal); Conceptualization (supporting). **Andrei L Kholkin**: Conceptualization (supporting). **Yulian M. Vysochanskii**: Conceptualization (supporting); Writing – review and editing (equal); Supervision.

**Acknowledgements.** A.N.M. and A.L.K. acknowledge support from the Horizon Europe Framework Programme (HORIZON-TMA-MSCA-SE), project № 101131229, Piezoelectricity in 2D-materials: materials, modeling, and applications (PIEZO 2D). A.N.M. also acknowledges funding from the National Academy of Sciences of Ukraine (grant N 4.8/23-п, 'Innovative materials and systems with magnetic and/or electrodipole ordering for the needs of using spintronics and nanoelectronics in strategically important issues of new technology). E.A.E. and L.P.Y. acknowledge support from the National Academy of Sciences of Ukraine. S.V.K. was supported by the center for 3D Ferroelectric Microelectronics (3DFeM), an Energy Frontier Research Center funded by the U.S. Department of Energy (DOE), Office of Science, Basic Energy Sciences under Award Number DE-SC0021118. This work (A.L.K.) was developed within the scope of project CICECO-Aveiro Institute of Materials (UIDB/50011/2020 & UIDP/50011/2020) financed by national funds through the FCT-Foundation for Science and Technology (Portugal).

**Data availability statement.** Numerical results presented in the work are obtained and visualized using a specialized software, Mathematica 13.2 [32]. The Mathematica notebook, which contain the codes, is available per reasonable request.

# Supplementary Materials to
# "The strain-induced transitions of the piezoelectric, pyroelectric and electrocaloric properties of the CuInP$_2$S$_6$ films"

### APPENDIX A. LGD parameters for a bulk ferroelectric CuInP$_2$S$_6$

**Table SI.** LGD parameters for a bulk ferrielectric CuInP$_2$S$_6$ at fixed stress determined in Refs.[33 - 34] from the experimental results for the dielectric permittivity [35, 36], spontaneous polarization [37], and lattice constants [38] as a function of hydrostatic pressure. Elastic compliances $s_{ij}$ were calculated from ultrasound velocity measurements [39, 40].

| Coefficient | Numerical value |
| --- | --- |
| $\varepsilon_b$ | 9 |
| $\alpha_T$ (C$^{-2}$·m J/K) | 1.64067×10$^7$ |
| $T_C$ (K) | 292.67 |
| $\beta$ (C$^{-4}$·m$^5$J) | 3.148×10$^{12}$ |
| $\gamma$ (C$^{-6}$·m$^9$J) | −1.0776×10$^{16}$ |
| $\delta$ (C$^{-8}$·m$^{13}$J) | 7.6318×10$^{18}$ |
| $Q_{i3}$ (C$^{-2}$·m$^4$) | $Q_{13} = 1.70136 - 0.00363\,T$<br>$Q_{23} = 1.13424 - 0.00242\,T$<br>$Q_{33} = -5.622 + 0.0105\,T$ |
| $Z_{i33}$ (C$^{-2}$·m$^4$) | $Z_{133} = -2059.65 + 0.8\,T$<br>$Z_{233} = -1211.26 + 0.45\,T$<br>$Z_{333} = 1381.37 - 12\,T$ |
| $s_{ij}$ (Pa$^{-1}$) | $s_{11} = 1.510\times10^{-11}$<br>$s_{12} = 0.183\times10^{-11}$ |

### APPENDIX B. Free energy with renormalized coefficients

Using results [41] and making the Legendre transformation of Eq.(1) to the strain-polarization representation, $\tilde{G} = G + u\sigma$, the renormalized free energy density is [34]:

$$\tilde{g}_{LD+ES} = \frac{\alpha^*}{2}P_3^2 + \frac{\beta^*}{4}P_3^4 + \frac{\gamma^*}{6}P_3^6 + \frac{\delta^*}{8}P_3^8 - P_3 E_3. \tag{S.1}$$

The coefficients $\alpha^*$, $\beta^*$, $\gamma^*$, and $\delta^*$ in Eq.(S.1) depend on the strain, $u_m$, and film thickness $h$ as:

$$\frac{\alpha^*}{2} = \frac{\alpha}{2} + \frac{1}{\varepsilon_0(\varepsilon_b + h/\lambda)} + \frac{u_m(Q_{23}(s_{12}-s_{11}) + Q_{13}(s_{12}-s_{22}))}{s_{11}s_{22}-s_{12}^2} - u_m^2 \frac{(s_{12}-s_{22})^2 W_{113} + (s_{11}-s_{12})^2 W_{223}}{2(s_{12}^2 - s_{11}s_{22})^2},$$

(S.2a)

$$\frac{\beta^*}{4} = \frac{\beta}{4} + \frac{Q_{23}^2 s_{11} - 2Q_{13}Q_{23}s_{12} + Q_{13}^2 s_{22}}{2(s_{11}s_{22} - s_{12}^2)} + u_m \left\{ \frac{(s_{22}-s_{12})Z_{133} + (s_{11}-s_{12})Z_{233}}{s_{12}^2 - s_{11}s_{22}} + \right.$$

$$\left. \frac{Q_{23}(s_{12}(s_{12}-s_{22})W_{113} + s_{11}(s_{11}-s_{12})W_{223}) + Q_{13}(s_{22}(-s_{12}+s_{22})W_{113} + s_{12}(-s_{11}+s_{12})W_{223})}{(s_{12}^2 - s_{11}s_{22})^2} \right\}, \quad \text{(S.2b)}$$

$$\frac{\gamma^*}{6} = \frac{\gamma}{6} + \frac{(Q_{13}s_{22} - Q_{23}s_{12})Z_{133} + (Q_{23}s_{11} - Q_{13}s_{12})Z_{233}}{s_{11}s_{22} - s_{12}^2} -$$

$$\frac{-2Q_{13}Q_{23}s_{12}(s_{22}W_{113} + s_{11}W_{223}) + Q_{23}^2(s_{12}^2 W_{113} + s_{11}^2 W_{223}) + Q_{13}^2(s_{22}^2 W_{113} + s_{12}^2 W_{223})}{2(s_{12}^2 - s_{11}s_{22})^2} +$$

$$u_m \frac{s_{22}^2 W_{113}Z_{133} + s_{11}^2 W_{223}Z_{233} - s_{12}(s_{22}W_{113} + s_{11}W_{223})(Z_{133}+Z_{233}) + s_{12}^2(W_{223}Z_{133} + W_{113}Z_{233})}{(s_{12}^2 - s_{11}s_{22})^2}, \quad \text{(S.2c)}$$

$$\frac{\delta^*}{8} = \frac{\delta}{8} + \frac{s_{22}Z_{133}^2 - 2s_{12}Z_{133}Z_{233} + s_{11}Z_{233}^2}{2(s_{11}s_{22} - s_{12}^2)} +$$

$$\frac{Q_{23}\left(s_{12}(s_{22}W_{113} + s_{11}W_{223})Z_{133} - s_{12}^2 W_{113}Z_{233} - s_{11}^2 W_{223}Z_{233}\right)}{(s_{12}^2 - s_{11}s_{22})^2} +$$

$$\frac{Q_{13}(-s_{22}^2 W_{113}Z_{133} + s_{12}s_{22}W_{113}Z_{233} + s_{12}W_{223}(-s_{12}Z_{133} + s_{11}Z_{233}))}{(s_{12}^2 - s_{11}s_{22})^2}. \quad \text{(S.2d)}$$

Since $W_{ijk}$ are small enough, we remain only linear terms in $W_{ijk}$ in Eqs.(S.2), and omit all terms proportional to higher powers of the parameter.

## References

[1]     Y. Liu, J. F. Scott, B. Dkhil. Direct and indirect measurements on electrocaloric effect: Recent developments and perspectives. Appl. Phys. Rev. **3**, 031102 (2016).

[2]     R. Kumar, S. Singh, Giant electrocaloric and energy storage performance of [(K$_{0.5}$Na$_{0.5}$)NbO$_3$](1−x) − [LiSbO$_3$]x nanocrystalline ceramics. Sci. Rep. **8**, 3186 (2018).

[3]     Zhao, J. Z., L. C. Chen, B. Xu, B. B. Zheng, J. Fan, and H. Xu. "Strain-tunable out-of-plane polarization in two-dimensional materials. Phys. Rev. B **101**, 121407 (2020).

[4]     Ju. Banys, A. Dziaugys, K. E. Glukhov, A. N. Morozovska, N. V. Morozovsky, Yu. M. Vysochanskii. Van der Waals Ferroelectrics: Properties and Device Applications of Phosphorous Chalcogenides (John Wiley & Sons, Weinheim 2022) 400 pp. ISBN: 978-3-527-35034-6

[5]     Zhi-Zheng Sun, Wei Xun, Li Jiang, Jia-Lin Zhong, and Yin-Zhong Wu. Strain engineering to facilitate the occurrence of 2D ferroelectricity in CuInP$_2$S$_6$ monolayer. Journal of Physics D: Applied Physics **52**, 465302 (2019).

[6]     Y.M. Vysochanskii, V.A. Stephanovich, A.A. Molnar, V.B. Cajipe, and X. Bourdon, Raman spectroscopy study of the ferrielectric-paraelectric transition in layered CuInP$_2$S$_6$. Phys. Rev. B, **58**, 9119 (1998).

[7]     N. Sivadas, P. Doak, and P. Ganesh. Anharmonic stabilization of ferrielectricity in CuInP$_2$Se$_6$. Phys. Rev. Research **4**, 013094 (2022).

[8]     J.A. Brehm, S.M. Neumayer, L. Tao, O'Hara, A., M. Chyasnavichus, M.A. Susner, M.A. McGuire, S.V. Kalinin, S. Jesse, P. Ganesh, S. Pantelides, P. Maksymovych and N. Balke. Tunable quadruple-well ferroelectric van der Waals crystals. Nat. Mater. **19**, 43 (2020).

[9]     A. N. Morozovska, E.A. Eliseev, M. E. Yelisieiev, Yu. M. Vysochanskii, and D. R. Evans. Stress-Induced Transformations of Polarization Switching in $CuInP_2S_6$ Nanoparticles. Physical Review Applied **19,** 054083 (2023), https://link.aps.org/doi/10.1103/PhysRevApplied.19.054083

[10]    A. N. Morozovska, E. A. Eliseev, A. Ghosh, M. E. Yelisieiev, Y. M. Vysochanskii, and S. V. Kalinin. Anomalous Polarization Reversal in Strained Thin Films of $CuInP_2S_6$, Phys. Rev. B, **108**, 054107 (2023) https://link.aps.org/doi/10.1103/PhysRevB.108.054107

[11]    A. N. Morozovska, E. A. Eliseev, Y. Liu, K. P. Kelley, A. Ghosh, Y. Liu, J. Yao, N. V. Morozovsky, A. L Kholkin, Y. M. Vysochanskii, and S. V. Kalinin. Bending-induced isostructural transitions in ultrathin layers of van der Waals ferrielectrics, https://doi.org/10.48550/arXiv.2305.15247

[12]    Y. Liu, A. N. Morozovska, A. Ghosh, K. P. Kelley, E. A. Eliseev, J. Yao, Y. Liu, and S. V. Kalinin, Disentangling stress and curvature effects in layered 2D ferroelectric $CuInP_2S_6$. https://doi.org/10.48550/arXiv.2305.14309

[13]    X. Bourdon, V. Maisonneuve, V.B. Cajipe, C. Payen, and J.E. Fischer. Copper sublattice ordering in layered $CuMP_2Se_6$ (M=In, Cr). J. All. Comp. **283**, 122 (1999).

[14]    V. Maisonneuve, V. B. Cajipe, A. Simon, R. Von Der Muhll, and J. Ravez. Ferrielectric ordering in lamellar $CuInP_2S_6$. Phys. Rev. B **56**, 10860 (1997).

[15]    N.A. Pertsev, A.G. Zembilgotov, A. K. Tagantsev, Effect of Mechanical Boundary Conditions on Phase Diagrams of Epitaxial Ferroelectric Thin Films, Phys. Rev. Lett. **80,** 1988 (1998).

[16]    C. Ederer, and N.A. Spaldin, Effect of epitaxial strain on the spontaneous polarization of thin film ferroelectrics, Phys. Rev. Lett. **95**, 257601 (2005).

[17]    A. Kvasov and A.K. Tagantsev. Role of high-order electromechanical coupling terms in thermodynamics of ferroelectric thin films. Phys. Rev. B **87**, 184101 (2013).

[18]    L. D. Landau, and I. M. Khalatnikov. On the anomalous absorption of sound near a second order phase transition point. In Dokl. Akad. Nauk SSSR **96**, 496 (1954).

[19]    Yu. M. Vysochanskii, M.M. Mayor, V. M. Rizak, V. Yu. Slivka, and M. M. Khoma. The tricritical Lifshitz point on the phase diagram of $Sn_2P_2(Se_xS_{1-x})_6$. Soviet Journal of Experimental and Theoretical Physics **95**, 1355 (1989).

[20]    A. Kohutych, R. Yevych, S. Perechinskii, V. Samulionis, J. Banys, and Yu. Vysochanskii. Sound behavior near the Lifshitz point in proper ferroelectrics. Phys. Rev. B **82**, 054101 (2010).

[21]    A. N. Morozovska, E. A. Eliseev, S. V. Kalinin, Y. M. Vysochanskii, and Petro Maksymovych. Stress-Induced Phase Transitions in Nanoscale $CuInP_2S_6$. Phys. Rev. B **104**, 054102 (2021).

[22]    P. Guranich, V.Shusta, E.Gerzanich, A.Slivka, I.Kuritsa, O.Gomonnai. "Influence of hydrostatic pressure on the dielectric properties of $CuInP_2S_6$ and $CuInP_2Se_6$ layered crystals." Journal of Physics: Conference Series **79**, 012009 (2007).

[23] A. V. Shusta, A. G. Slivka, V. M. Kedylich, P. P. Guranich, V. S. Shusta, E. I. Gerzanich, I. P. Prits, Effect of uniaxial pressure on dielectric properties of CuInP$_2$S$_6$ crystals. Scientific Bulletin of Uzhhorod University. Physical series, **28**, 44 (2010).

[24] A. Kohutych, V. Liubachko, V. Hryts, Yu. Shiposh, M. Kundria, M. Medulych, K. Glukhov, R. Yevych, and Yu. Vysochanskii. Phonon spectra and phase transitions in van der Waals ferroics MM'P$_2$X$_6$, Molecular Crystals and Liquid Crystals **747**, 14 (2022). https://doi.org/10.1080/15421406.2022.2066787

[25] M. A. Susner, M. Chyasnavichyus, A. A. Puretzky, Q. He, B. S. Conner, Y. Ren, D. A. Cullen et al. Cation–Eutectic Transition via Sublattice Melting in CuInP$_2$S$_6$/In$_{4/3}$P$_2$S$_6$ van der Waals Layered Crystals. ACS nano, **11**, 7060 (2017).

[26] J. Banys, J. Macutkevic, V. Samulionis, A. Brilingas & Yu. Vysochanskii, Dielectric and ultrasonic investigation of phase transition in CuInP$_2$S$_6$ crystals. Phase Transitions: A Multinational Journal **77**, 345 (2004).

[27] V. Samulionis, J. Banys, Yu. Vysochanskii, and V. Cajipe. Elastic and electromechanical properties of new ferroelectric-semiconductor materials of Sn$_2$P$_2$S$_6$ family. Ferroelectrics **257**, 113 (2001).

[28] A. N. Morozovska, E. A. Eliseev, M. D. Glinchuk, H. V. Shevliakova, G. S. Svechnikov, M. V. Silibin, A. V. Sysa, A. D. Yaremkevich, N. V. Morozovsky, and V. V. Shvartsman. Analytical description of the size effect on pyroelectric and electrocaloric properties of ferroelectric nanoparticles. Phys. Rev. Materials **3**, 104414 (2019). https://link.aps.org/doi/10.1103/PhysRevMaterials.3.104414

[29] F. Jona, and G. Shirane. Ferroelectric Crystals, International Series of Monographs on Solid State Physics. Pergamon press (1962).

[30] K. Moriya, N. Kariya, A. Inaba, T. Matsuo, I. Pritz, and Y.M. Vysochanskii. "Low-temperature calorimetric study of phase transitions in CuCrP2S6." Solid state communications **136**, 173 (2005). https://doi.org/10.1016/j.ssc.2005.05.040

[31] K. Moriya, N. Kariya, I. Pritz, Yu M. Vysochanskii, A. Inaba, T. Matsuo, The Heat Capacity of CuInP2Se6, a Layer-Structured Selenodiphosphate. Ferroelectrics **346**, 143 (2007). https://doi.org/10.1080/00150190601182402

[32] https://www.wolfram.com/mathematica

[33] A. N. Morozovska, E.A. Eliseev, M. E. Yelisieiev, Yu. M. Vysochanskii, and D. R. Evans. Stress-Induced Transformations of Polarization Switching in CuInP$_2$S$_6$ Nanoparticles. Physical Review Applied **19,** 054083 (2023), https://link.aps.org/doi/10.1103/PhysRevApplied.19.054083

[34] A. N. Morozovska, E. A. Eliseev, A. Ghosh, M. E. Yelisieiev, Y. M. Vysochanskii, and S. V. Kalinin. Anomalous Polarization Reversal in Strained Thin Films of CuInP$_2$S$_6$, Phys. Rev. B, **108**, 054107 (2023) https://link.aps.org/doi/10.1103/PhysRevB.108.054107

[35]     P. Guranich, V.Shusta, E.Gerzanich , A.Slivka, I.Kuritsa, O.Gomonnai. "Influence of hydrostatic pressure on the dielectric properties of $CuInP_2S_6$ and $CuInP_2Se_6$ layered crystals." Journal of Physics: Conference Series **79**, 012009 (2007).

[36]     A. V. Shusta, A. G. Slivka, V. M. Kedylich, P. P. Guranich, V. S. Shusta, E. I. Gerzanich, I. P. Prits, Effect of uniaxial pressure on dielectric properties of $CuInP_2S_6$ crystals. Scientific Bulletin of Uzhhorod University. Physical series, **28**, 44 (2010).

[37]     A. Kohutych, V. Liubachko, V. Hryts, Yu. Shiposh, M. Kundria, M. Medulych, K. Glukhov, R. Yevych, and Yu. Vysochanskii. Phonon spectra and phase transitions in van der Waals ferroics MM'$P_2X_6$, Molecular Crystals and Liquid Crystals **747**, 14 (2022). https://doi.org/10.1080/15421406.2022.2066787

[38]     M. A. Susner, M. Chyasnavichyus, A. A. Puretzky, Q. He, B. S. Conner, Y. Ren, D. A. Cullen et al. Cation–Eutectic Transition via Sublattice Melting in $CuInP_2S_6$/$In_{4/3}P_2S_6$ van der Waals Layered Crystals. ACS nano, **11**, 7060 (2017).

[39]     J. Banys, J. Macutkevic, V. Samulionis, A. Brilingas & Yu. Vysochanskii, Dielectric and ultrasonic investigation of phase transition in $CuInP_2S_6$ crystals. Phase Transitions: A Multinational Journal **77**, 345 (2004).

[40]     V. Samulionis, J. Banys, Yu. Vysochanskii, and V. Cajipe. Elastic and electromechanical properties of new ferroelectric-semiconductor materials of $Sn_2P_2S_6$ family. Ferroelectrics **257**, 113 (2001).

[41]     A. Kvasov and A.K. Tagantsev. Role of high-order electromechanical coupling terms in thermodynamics of ferroelectric thin films. Phys. Rev. B **87**, 184101 (2013).